# The diamond Nitrogen-Vacancy center as a probe of random fluctuations in a nuclear spin ensemble

Abdelghani Laraoui[1], Jonathan S. Hodges[2], Colm Ryan[3], and Carlos A. Meriles[1,*]

[1]Department of Physics, City College of New York – CUNY, New York, NY 10031, USA

[2]Department of Electrical Engineering, Columbia University, New York, NY 10023, USA

[3]Department of Nuclear Science and Engineering, Massachusetts Institute of Technology, Cambridge, MA 02139, USA

**Abstract**

New schemes that exploit the unique properties of Nitrogen-Vacancy (NV) centers in diamond are presently being explored as a platform for high-resolution magnetic sensing. Here we focus on the ability of a NV center to monitor an adjacent mesoscopic nuclear spin bath. For this purpose, we conduct comparative experiments where the NV spin evolves under the influence of surrounding $^{13}$C nuclei or, alternatively, in the presence of asynchronous AC fields engineered to emulate bath fluctuations. Our study reveals substantial differences that underscore the limitations of the semi-classical picture when interpreting and predicting the outcome of experiments designed to probe small nuclear spin ensembles. In particular, our study elucidates the NV center response to bath fluctuations under common pulse sequences, and explores a detection protocol designed to probe time correlations of the nuclear spin bath dynamics. Further, we show that the presence of macroscopic nuclear spin order is key to the emergence of semi-classical spin magnetometry.

---

[*] To whom correspondence should be addressed. E-mail: cmeriles@sci.ccny.cuny.edu

**I. Introduction**

Nitrogen-Vacancy (NV) centers in diamond are presently the focus of a broad cross-disciplinary research effort combining the fields of condensed matter, atomic physics, and precision metrology. Several unique properties, including the long spin coherence times and superb photostability at room temperature, make them central to various proposals for quantum information processing[1] and quantum cryptography.[2] Key to the present study is the use of NVs for high-resolution magnetometry, an application facilitated by the ability to initialize, manipulate, and readout spins with high fidelity.[3] Initial experimental work has used single NVs to map the magnetic field created by a ferromagnetic particle with nanoscale resolution,[4] and has demonstrated detection of synchronous, coil-induced AC fields[5] with sensitivity approaching 30 nT/Hz$^{1/2}$. Further, arrays of superficial NVs have been exploited to determine the local amplitude *and* direction of the magnetic field generated by current-carrying wire strips imprinted on the host diamond surface.[6-9]

While the use of single NVs to control and probe few adjacent, strongly-coupled nuclear spins is well documented,[10] extensions that target the monitoring of more numerous, weakly-coupled nuclear ensembles are still pending. Motivating this line of research are applications where the NV is envisioned as a high-resolution *nuclear spin* magnetometer capable of providing information on the local density, chemical composition, physical structure and/or dynamical processes within 'mesoscopic' (~10 nm) regions of an 'a-priori-unknown' sample. In an initial theoretical approach to the problem[11], a mesoscopic, unpolarized nuclear spin ensemble at room temperature was modeled via a stochastic, time-dependent magnetic field interacting with a single NV center. The resulting description is 'semi-classical' as the 'back-action' of the nuclear spin ensemble, which generates NV-nuclei entanglement, is not considered.

Here we revisit this topic by observing a single NV center alternatively exposed to the surrounding 'bath' of $^{13}$C spins or an external, coil-generated magnetic field. Despite the ability of the semi-classical approach to explain the observed NV response to common pulse sequences (e.g., excitation-evolution, Hahn-echo, etc), our results expose fundamental limitations in its ability to predict the outcome of an arbitrary nuclear spin sensing protocol. We illustrate the problem through an experiment designed to probe the slow, long-term correlation function of the effective nuclear magnetic field acting on the NV center. Moreover, we tackle some common misconceptions when interpreting the dynamics of an NV center in a fluctuating nuclear spin bath, derive analytical expressions that clarify on the role of adjacent (versus distant) nuclear spins, and investigate the conditions necessary to regain the limit of semiclassical magnetometry.

Our manuscript is organized in the following way: Section II briefly touches on some practical details, explains the pulse protocols used, and presents our experimental observations. Through quantum mechanical calculations and 'disjoint cluster' simulations of the combined nuclear ensemble-NV system, Section III discusses the NV response in the presence of a slowly fluctuating nuclear 'bath'. Finally, concluding remarks and technical details on some of the analytical formulae and numerical simulations are presented in Section IV and the Appendices, respectively.

## II. Experimental Protocols

II.a Experimental Methods

The theory and practice of single NV center detection and control is well described in the literature and will be only briefly reviewed. The negatively charged NV center addressed here comprises a substitutional Nitrogen associated with a vacant, adjacent lattice site. The ground state is a spin triplet with a concomitant 'crystal field' aligned along the [111] axis (or its crystallographic equivalents). The zero-field (or crystal field) splitting between magnetic sublevels $|m_S = 0\rangle$ and $|m_S = \pm 1\rangle$ is Δ=2.87 GHz. Green light illumination produces, to a good approximation, a spin conserving transition to the first excited state — also a triplet — which, in turn, leads to broadband photoluminescence emission. Non-radiative inter-system crossing transitions to lower-energy metastable states are strongly spin selective as the shelving rate from $|m_S = 0\rangle$ is much smaller than those from $|m_S = \pm 1\rangle$, thus allowing for an optical readout of the spin state (i.e., the $|m_S = 0\rangle$ state is brighter). The spin-selective crossing rates also lead to an almost complete pumping of the spin degree of freedom after ~1 μs illumination.[12]

Experiments are carried out in a high-purity diamond crystal with NV concentration lower than 10 ppb. We optically address individual centers using a purpose-built confocal microscope. Electron spin resonance (ESR) spectroscopy of the spin sublevels is carried out using a thin copper wire (20 μm) overlaid on the crystal surface. We apply a weak, DC magnetic field $B_0$ (~4 mT) collinear with the crystal field to break the degeneracy between the $|m_S = \pm 1\rangle$ states. This allows us to selectively address one of the two possible transitions (e.g., $|m_S = 0\rangle \rightarrow |m_S = 1\rangle$), which virtually renders the NV a two-level system. When necessary, we use a four-turn coil in the vicinity of the host diamond to generate an auxiliary field $b_A(t) = b_{A\_0} + b_{A\_1} \cos(\omega_A t + \varepsilon_A)$; a computer-controlled wavefunction generator is used to set amplitudes $b_{A\_0}$, $b_{A\_1}$, frequency $\omega_A$ and phase $\varepsilon_A$. The system is designed to introduce random amplitude and phase changes over a time scale of a few milliseconds, longer than the time necessary for a single NV observation (<200 μs) but much shorter than the averaging time (~ 1 minute) required to overcome the shot noise.

II.b. Semi-Classical Response of the NV Center to Echo and Ramsey Sequences

To investigate the effect of a nuclear spin ensemble on the NV signal, we start with the response of a typical color center to the Hahn-echo sequence $(\pi/2)_y$-$\tau$-$(\pi)_y$-$\tau$-$(\pi/2)_y$ shown in Fig. 1a. As reported in prior observations,[13] the signal exhibits periodic 'collapses and revivals' whose amplitude progressively decays with a time constant that approaches the NV 'true' coherence time $T_2$. The time separation between consecutive revivals coincides with the $^{13}$C Larmor frequency at the applied magnetic field (4 mT in the present case), a key feature that motivates a simple, heuristic interpretation: As the nuclear spins surrounding the NV center precess around $B_0$ with frequency $\omega_N$ they effectively create a random magnetic field $b_N(t) = b_{N\_0} + b_{N\_1} \cos(\omega_N t + \varepsilon_N)$, whose amplitude and phase slowly change with time. Then, the phase picked-up by the NV center during the first half of the Hahn-echo sequence differs from that accumulated during the second half and depends on the relative phase of the nuclear field, thus resulting in a signal 'collapse' after a time-ensemble average. If, however, the nuclear spins have enough time to complete one full rotation during $\tau$, the NV accumulates no net phase and an echo 'revival' takes place.

As shown in Fig. 1c, one can use the coil-generated field $b_A(t)$ to induce an identical response. This response is more generic than an artificial situation: It could also, for example, come from physical motion of the NV center in a static gradient field. In order to eliminate the influence of $^{13}$C spins on the signal modulation, we progressively adjust the DC field $B_0$ (see upper axis in Fig. 1b) so that the nuclear Larmor frequency $\omega_N/2\pi = (\gamma_N B_0)/2\pi$ exactly matches the inverse of a desired evolution time $\tau$ (i.e., nuclear spins undergo one full cycle at every given $\tau$). The result is a train of $b_A$-induced collapses and revivals at the frequency $\omega_A$.

We now find an analytical expression that formally describes these ideas in the limit of infinitely short pulses. By detuning the $|m_s = -1\rangle \rightarrow |m_s = 0\rangle$ transition from our microwave field, we can treat the NV spin as a two-level system. Within the $m_S = \{0, 1\}$ sub-manifold, we

thus define the normalized NV signal

$$S = \langle Tr\{\rho\sigma_z\}\rangle, \qquad (1)$$

where $\rho$ denotes the system density matrix and $\sigma_z = |0\rangle\langle 0| - |1\rangle\langle 1|$ is the Pauli operator in the direction of the crystal field (assumed along the z-axis); brackets indicate time (or ensemble) average (see below). Within a time interval shorter than $T_2$, the NV response to a Hahn-echo sequence in the presence of a 'classical' field $b_A(t)$ is given by

$$S_C^{HE}(\tau) = \langle \cos\phi_{12}\rangle, \qquad (2)$$

where $\phi_n$, $n=1,2$ denotes the phase picked up by the NV center during the $n$-th inter-pulse interval and $\phi_{12} \equiv \phi_1 - \phi_2 = (4\gamma_{NV} b_{A\_1}/\omega_A)\sin^2(\omega_A\tau/2)\sin(\omega_A\tau+\varepsilon_A)$; $\gamma_{NV}$ is the color center gyromagnetic ratio. In the limit $\phi_n \ll 1$, Eq. (2) yields

$$\begin{aligned}S_C^{HE}(\tau) &\cong 1 - 8\langle(\gamma_{NV} b_{A\_1}/\omega_A)^2 \sin^2(\omega_A\tau+\varepsilon_A)\rangle\sin^4(\omega_A\tau/2) \\ &= 1 - 4(\gamma_{NV}^2 \langle b_{A\_1}^2\rangle/\omega_A^2)\sin^4(\omega_A\tau/2),\end{aligned} \qquad (3)$$

where the last expression holds for the case in which phase and amplitude fluctuations are independent and the phase distribution function is uniform over $[0,2\pi]$. Eq. (3) predicts revivals at times $\tau_m = 2\pi m/\omega_A$ with $m$ integer. As expected, the signal is immune to the DC field component $b_{A\_0}$, but the depth of the modulation (or, in the more general case, the duration of the revival) can be controlled via $b_{A\_1}$.

For future reference, we note that the present picture can be easily extended to a Ramsey sequence $(\pi/2)_y$-$t$-$(\pi/2)_y$: In this case, one imagines the nuclear field $b_{N\_0}$ (or, correspondingly, $b_{A\_0}$ in the coil-induced analog) changing randomly over time, with the concomitant signal interference and monotonic decay over a time $T_2^*$ when averaged in a time ensemble. Analytically, and in the same approximation as in (3), we find

$$S_C^R(t) = \langle \cos\phi \rangle \cong 1 - \left(\gamma_{NV}^2 \langle b_{A\_1}^2 \rangle / \omega_A^2\right)\sin^2(\omega_A \tau/2) - \left(\gamma_{NV}^2 \langle b_{A\_0}^2 \rangle t^2 / 2\right), \tag{4}$$

where, as before, we assumed $1 \gg \phi = (2\gamma_{NV} b_{A\_1}/\omega_A)\sin(\omega_A t/2)\cos(\omega_A t/2 + \varepsilon_A) + \gamma_{NV} b_{A\_0} t$ and $t < T_2^*$.

II.c. Probing the Noise Correlation Time

In line with recent experiments designed to probe the spin noise present in a small ensemble[14], the above framework suggests that it should be possible to use the NV center to determine the average amplitude, central frequency, and time correlation of the acting random field. While alternate routes are conceivable[8], Fig. 2 shows a flexible approach comprising two echo-sequences of fixed, identical duration $2\tau$ separated by a variable free evolution time $\tilde{\tau}$. This sequence shares some similarities and motivation with 2D exchange experiments[15] such as NOESY or hyperfine correlation protocols[16] such as HYSCORE or DEFENCE. In the presence of a random magnetic field $b_A(t)$, and with proper phase cycling this yields a signal (see Appendix A)

$$S_C^{Cor}(\tau, \tilde{\tau}) = \langle \sin\phi_{12} \sin\phi_{45} \rangle. \tag{5}$$

Similar to (2), $\phi_n$, $n=1...5$ denotes the accumulated phase during the $n$-th free-evolution interval and $\phi_{n,n+1} \equiv \phi_n - \phi_{n+1}$. Eq. (5) indicates that the present sequence can be used as a tool to characterize the time coherence of the field under investigation (so long as the field correlation time $\tilde{\tau}_c$ is shorter than the NV longitudinal relaxation time $T_1$). For the particular setting $\omega_A(2\tau + \tilde{\tau}) = 2m\pi$ with $m$ integer, and in the case $\tilde{\tau} < \tilde{\tau}_c, T_1$ we obtain for random $\varepsilon_A$ but *fixed* $b_{A\_1}$

$$S_C^{Cor}(\tau, \tilde{\tau}) = \frac{1}{2}(1 - J_0(2K)), \tag{6}$$

where $J_0$ denotes the zero-order Bessel function and $K \equiv (4\gamma_{NV} b_{A\_1}/\omega_A)\sin^2(\omega_A \tau/2)$. Fig. 2

demonstrates reasonable agreement between observed and predicted responses as a function of $b_{A\_1}$ (Fig. 2b) and $\tilde{\tau}$ (Figs. 2c-2e). In the more general case of variable $b_{A\_1}$ and $\omega_A(2\tau+\tilde{\tau}) \neq 2m\pi$ an expression can be attained when $\phi_n \ll 1$. One finds

$$S_C^{Cor}(\tau,\tilde{\tau}) \cong \langle \phi_{12}\phi_{45}\rangle \sim \left(16\gamma_{NV}^2 \langle b_{A\_1}^2\rangle / \omega_A^2\right) \sin^4(\omega_A \tau/2)\cos(\omega_A(2\tau+\tilde{\tau})), \quad (7)$$

where, once again, we assumed $\tilde{\tau} < \tilde{\tau}_c, T_1$.

With a natural abundance of only ~1%, $^{13}$C spins experience relatively weak homonuclear dipolar couplings of order ~100 Hz implying that if these were the only non-commuting dynamics, once formed, random coherences should persist for several milliseconds. In the spirit of the experiment of Fig. 1, one should thus be able to anticipate the effect of the $^{13}$C bath in the above sequence with our engineered AC field $b_A(t)$. We do this in Fig. 3a for a root mean square (rms) field amplitude $\sqrt{\langle b_{A\_1}^2\rangle}$ of ~4 μT as determined from a comparison between the coil- and $^{13}$C-induced echo signals of Fig. 1. In good agreement with our calculations, we find a strong correlation signal whose period coincides with $\Delta\tilde{\tau} = 2\pi/\omega_A$ as predicted by (7). We note, however, that in this case $4\gamma_{NV}\sqrt{\langle b_{A\_1}^2\rangle}/\omega_A \gg 1$ largely exceeds the NV linear range ($b_{A\_1}$<300 nT in Fig. 2b). This causes interference between sinusoidal (Fig. 2c) and rapidly oscillating responses (Figs. 2d-2e) and leads to a pattern of sharp positive (and negative) peaks correctly reproduced by the numerical computation of Eq. (5) (solid line).

In stark contrast with these observations, however, Fig. 3b shows that no distinguishable signal is present when $^{13}$C spins are the source of the fluctuating field, a result we confirmed with observations in several, distinct NV centers. This response (or, rather, the lack thereof) is intriguing, particularly given the short time scale of the experiment (where $\tilde{\tau} \sim 2\tau$ and $(4\tau+\tilde{\tau}) \ll T_2$). While one can argue that the effective nuclear field (and, therefore, the amplitude of the resulting modulation pattern) could possibly be smaller than that assumed, we emphasize that $\sqrt{\langle b_{A\_1}^2 \rangle}$ ~4 µT largely 'saturates' the NV response (see Fig. 2b). With the present signal-to-noise ratio, one can thus rule out fluctuating fields down to ~100 nT, which, in turn, is much too small a value to reproduce the collapses and revivals of Fig. 1 (see dashed, fainted line in Fig 1b). These observations point to an incomplete description of the relevant NV-nuclear spin dynamics under a semi-classical model.

**III. Discussion**

III.a. Quantum Response of the NV center

To more thoroughly describe the problem we develop a fully quantum mechanical formulation of the dynamics. We start by defining a Hamiltonian that explicitly takes into account the presence of *M* surrounding nuclear spins. Taking a secular approximation with the zero-field splitting defining the quantization axis, we write

$$H = \Delta S_z^2 - \gamma_{NV} S_z B_0 + \sum_{\substack{j=1 \\ q=x,y,z}}^{M} S_z A_{zq}^{(j)} I_q^{(j)} + \omega_{N\_0} \sum_{j=1}^{M} I_z^{(j)} + H_N'. \quad (8)$$

The first two terms denote the NV crystal field and Zeeman contributions while the third term expresses the coupling between the NV center and neighboring nuclear spins $I^{(j)}$. Nuclear Zeeman and spin-spin interactions are included through the fourth and fifth terms, respectively. Consistent with proposed sensing applications where the NV is only 'weakly' coupled to the system under consideration, we assume $A_{zq} \leq \omega_{N\_0}$. This simplification may indeed apply to the $^{13}$C environment of a small (but non-negligible) fraction of single NVs in our sample (e.g., for $B_0$~4 mT this condition is already met if the nearest $^{13}$C spin is at a distance of 0.8 nm or greater; higher fields reduce this distance further). Also, we limit our discussion to times much shorter than $T_2$, which allows us to ignore homonuclear dipolar interactions (i.e., we assume $H'_N \sim 0$).

Starting from Eq. (1) and after initializing the system to $|m_S = 0\rangle\langle m_S = 0| \otimes \rho_N$, one can show that for an arbitrary pulse sequence of the form $(\pi/2)_y$-$t_1$-pulse1-$t_2$-pulse2-…-$t_m$-$(\pi/2)_{-y}$ the signal is given by

$$S_Q(t) = \left\langle \text{Re}\left\{ \prod_{j=1}^{M} Tr_r \left\{ \rho_N^{(j)} \left[ (U_T^\dagger)_{10}(U_T)_{10} + (U_T^\dagger)_{00}(U_T)_{11} + (U_T^\dagger)_{00}(U_T)_{10} + (U_T^\dagger)_{10}(U_T)_{11} \right] \right\} \right\} \right\rangle, \quad (9)$$

where $Tr_r$ indicates reduced trace over nuclear spin states, $U_T$ is the time evolution operator (excluding projection and excitation pulses), and $(U_T)_{\alpha\beta} \equiv \langle m_S = \alpha | U_T | m_S = \beta \rangle$ with $\alpha, \beta = \{0, 1\}$. Eq. (9) assumes that all microwave pulses act selectively within the $m_S=\{0,1\}$ manifold and that $\rho_N$ can be expressed as the tensor product $\rho_N^{(1)} \otimes \rho_N^{(2)} \otimes \rho_N^{(3)} ... \otimes \rho_N^{(M)}$ of individual nuclear spins (a condition we will revisit later in the manuscript).

For the particular case of the Hahn-echo sequence, we have $U_T = U(\tau)R_y(\pi)U(\tau)$, where $R_y(\pi)$ denotes the $\pi$-rotation operator along the $y$-axis and $U(\tau)$ is the free evolution operator during the interpulse interval $\tau$. Eq. (9) then yields

$$S_Q^{HE}(\tau) = \left\langle \text{Re}\left\{ \prod_{j=1}^{M} Tr_r \left\{ \rho_N^{(j)} (U^\dagger)_{00} (U^\dagger)_{11} (U)_{00} (U)_{11} \right\} \right\} \right\rangle. \quad (10)$$

To model random fluctuations in the spin-1/2 nuclear bath, we write $\rho_N^{(j)}$ in the form $\rho_N^{(j)} = 1/2 + 2P_N^{(j)} I_z^{(j)} + 2T_N^{(j)} I_x^{(j)} \cos\varepsilon_j + 2T_N^{(j)} I_y^{(j)} \sin\varepsilon_j$, where $P_N^{(j)}$, $T_N^{(j)}$, and $\varepsilon_j$ are stochastic, independent parameters that characterize the polarization and transverse coherence of individual nuclei at the beginning of the echo sequence. After some algebra, we find (see Appendix B)

$$S_Q^{HE}(\tau) \cong \left\langle 1 - 2\sin^2(\omega_{N\_0}\tau/2) \left\{ \sum_{j=1}^{M} \tilde{\theta}_j^2 \sin^2(\omega_{N\_1}^{(j)}\tau/2) + 4\sum_{\substack{j,k=1 \\ j \neq k}}^{M} T_N^{(j)} T_N^{(k)} \tilde{\theta}_j \tilde{\theta}_k \sin(\omega_{N\_1}^{(j)}\tau/2)\sin(\omega_{N\_1}^{(k)}\tau/2) \times \right. \right.$$

$$\left. \left. \times \sin\left((\omega_{N\_0} + \omega_{N\_1}^{(j)})\tau/2 - \tilde{\varphi}_j - \varepsilon_j\right) \sin\left((\omega_{N\_0} + \omega_{N\_1}^{(k)})\tau/2 - \tilde{\varphi}_k - \varepsilon_k\right) \right\} \right\rangle, \quad (11)$$

where we define $\vec{\omega}_{N\_1}^{(j)} \equiv (A_{zx}^{(j)}, A_{zy}^{(j)}, A_{zz}^{(j)} + \omega_{N\_0})$, $\tan\tilde{\theta}_j \equiv \left((\omega_{N\_1,x}^{(j)})^2 + (\omega_{N\_1,y}^{(j)})^2\right)^{1/2} / \omega_{N\_1,z}^{(j)}$ and $\tan\tilde{\varphi}_j \equiv \omega_{N\_1,y}^{(j)} / \omega_{N\_1,x}^{(j)}$. Furthermore, we assume the nuclear Zeeman interaction is much stronger than the transverse hyperfine fields, such that $\tilde{\theta}_j \ll 1$ for all nuclear spins $j$. After ensemble average, the second sum cancels and Eq. (11) reduces to

$$S_Q^{HE}(\tau) \cong 1 - 2\sin^2(\omega_{N\_0}\tau/2) \sum_{j=1}^{M} \tilde{\theta}_j^2 \sin^2(\omega_{N\_1}^{(j)}\tau/2)$$
$$\approx 1 - 2\sin^4(\omega_{N\_0}\tau/2) \sum_{j=1}^{M} \tilde{\theta}_j^2. \quad (12)$$

To facilitate comparison with the semiclassical result, we express $b_{N\_1}$ in Eq. (3) as the superposition of contributions from fictitious, classical magnetic moments of transverse amplitude $T_N^{(j)}$. We start by expressing the field in the form

$$\gamma_{NV} b_{N\_1}(t) = \sum_{j=1}^{M} T_N^{(j)} \left( A_{zx}^{(j)} \cos(\omega_{N\_0} t + \varepsilon_j) + A_{zy}^{(j)} \sin(\omega_{N\_0} t + \varepsilon_j) \right)$$
$$\cong \omega_{N\_0} \sum_{j=1}^{M} T_N^{(j)} \tilde{\theta}_j \cos(\omega_{N\_0} t + \varepsilon_j - \tilde{\varphi}_j), \quad (13)$$

which allows us to calculate the accumulated phase (see Eq. (2)). Using the same approximation as in (3) and after taking ensemble average we find

$$S_C^{HE}(\tau) \cong \left\langle 1 - 8\sin^4(\omega_{N\_0}\tau/2) \left( \sum_{j=1}^{M} T_N^{(j)} \tilde{\theta}_j \sin(\omega_{N\_0}\tau + \varepsilon_j - \tilde{\varphi}_j) \right)^2 \right\rangle$$
$$= 1 - 4\sin^4(\omega_{N\_0}\tau/2) \sum_{j=1}^{M} \left\langle (T_N^{(j)})^2 \right\rangle \tilde{\theta}_j^2. \quad (14)$$

Interestingly, Eq. (14) predicts a pattern of collapses and revivals virtually identical to the quantum mechanical formula (Eq. (12)). Yet, the underlying physics is fundamentally different: While the echo modulations present in Eq. (14) depend on nuclear spins having a non-zero transverse polarization, the quantum expression (Eq. (12)) is completely insensitive to the exact state of the bath at the time of the experiment. Formally, the origin of the modulation arises from Eq. (10), which compares evolution under $U_{00}$ followed by $U_{11}$ with evolution in the inverse order. With an anisotropic hyperfine interaction these two operators do not, in general, commute, thus leading to a decay of the echo signal. However, if $\omega_{N\_0}\tau = 2m\pi$ then $U_{00} = \exp(2m\pi i I_z^{(j)}) = 1$ for all nuclei, and a revival takes place *regardless to the explicit form of* $\rho_N$. We therefore conclude that collapses and revivals are not only a consequence of classical precession of the statistical polarization of the bath. Echo modulations persist even if the state of surrounding nuclear spins is known. An important corollary is that echo collapses and revivals must be present even when neighboring nuclei are perfectly polarized before the Hahn-echo sequence is applied.

III.b. The Impact of the $^{13}$C Distribution

While the above discussion indicates that collapses and revivals occur for any separable initial nuclear spin configuration, a question relevant in nuclear magnetometry experiments concerns the role played by adjacent (versus distant) spins in an ensemble. Although it is true that dipolar interactions with *individual* spins decay as the inverse cube of the distance $r$ to the center, the number of nuclei between $r$ and $r+dr$ typically increases quadratically. The combined effect, well known in Nuclear Magnetic Resonance, gives rise to slowly-decaying, 'long-range' dipolar fields, which have proven useful to couple nuclear spins over macroscopic distances.[17-20] To more precisely identify the location of nuclei contributing the most to the echo modulations

in Eq. (12), we write

$$\sum_{j=1}^{M} \tilde{\theta}_j^2 \rightarrow \int_{\substack{Nuclear \\ Ensemble}} d^3r\, \kappa(\vec{r}) \frac{(A_{zx}^2 + A_{zy}^2)}{\omega_{N\_0}^2} \approx \frac{2\pi\kappa C^2}{3r_{min}^3}. \quad (15)$$

The last expression in (15) assumes a uniform nuclear spin density $\kappa(\vec{r}) = \kappa$ and uses $A_{zx}/\omega_{N\_0} = C\sin(2\theta)\cos\varphi/r^3$ and $A_{zy}/\omega_{N\_0} = C\sin(2\theta)\sin\varphi/r^3$ with the proportionality constant $C \equiv -(3\mu_0 \gamma_{NV} \gamma_N \hbar/8\pi\omega_{N\_0})$. As usual, $\mu_0$ is the vacuum magnetic permeability, $\gamma_N$ denotes the nuclear gyromagnetic ratio, $\hbar$ is Plank's constant divided by $2\pi$, and $\theta$ and $\varphi$ are the polar and azimuthal angles formed by the nuclear-NV vector in the laboratory reference frame (with the z-axis collinear with the crystal field); $r_{min}$ denotes the radius of the shell containing the nearest non-zero-spin nuclei. After replacing in (12) we find

$$S_Q^{HE}(\tau) \sim 1 - \frac{4\pi\kappa C^2}{3r_{min}^3} \sin^4(\omega_{N\_0}\tau/2). \quad (16)$$

Eq. (16) exposes a rapid fall-off of the depth of the modulation with distance: For example, for natural abundance $^{13}C$ spins in the diamond lattice and at the same 4 mT field, we get $4\pi\kappa C^2/(3r_{min}^3) \sim 1$ for $r_{min} \sim 2$ nm; these modulations, however, virtually decrease by an order of magnitude if $r_{min}$ doubles, implying that relevant nuclear spins lie within a small ~5-nm-radius sphere around the NV center. On a related comment, we mention that in an experiment where a ~20-nm-diameter diamond nanostructure probes an organic sample, $4\pi\kappa C^2/(3r_{min}^3)$ is of order $10^{-2}$ indicating that the interaction with surrounding protons will not lead to a detectable pattern of collapses and revivals.[11] We also note that the effective $r_{min}$ can be scaled by adjusting $B_0$; higher fields can make the effective $r_{min}$ small enough to suppress these effects for the majority of NV centers in natural abundance $^{13}C$ diamond.

III.c. The impact of $^{13}C$ Correlations

We return to Eq. (11), rewritten below for presentation purposes in the form

$$S_Q^{HE}(\tau) = 1 - 2\sin^4(\omega_{N\_0}\tau/2)\sum_{j=1}^{M}\tilde{\theta}_j^2 -$$

$$-8\sin^4(\omega_{N\_0}\tau/2)\left\langle\left(\sum_{j=1}^{M}T_N^{(j)}\tilde{\theta}_j\sin(\omega_{N\_0}\tau-\tilde{\varphi}_j-\varepsilon_j)\right)^2 - \sum_{j=1}^{M}\left(T_N^{(j)}\tilde{\theta}_j\sin(\omega_{N\_0}\tau-\tilde{\varphi}_j-\varepsilon_j)\right)^2\right\rangle, \quad (17)$$

As before, we used the approximation $\omega_{N\_1}^{(j)} \sim \omega_{N\_0}$. Comparison with Eqs. (13) and (14) shows that the second sum (third term) can be interpreted as resulting from a 'classical' nuclear spin field. To examine how nuclear spin order arises, consider the situation where the nuclear spins exhibit correlations relative to each other while remaining overall asynchronous with respect to the experimental repeat timing. At a given time, suppose the parameters $\varepsilon_j$ are chosen so as to generate the maximum possible dipolar field at the NV site. Here we choose a near-optimum configuration where $\langle T_N^{(j)}\rangle = T_N$ for all $j$, and $\varepsilon_j$ satisfies $\varepsilon_j(\theta,\varphi) = -(\varphi + \pi f(\theta) + \varepsilon_0)$ with $f(\theta)=0$ if $\theta \leq \pi/2$ and $f(\theta)=1$ otherwise (i.e., the phase grows with the azimuthal angle, and spins at opposite sides of the equator point in opposite directions); $\varepsilon_0$ is a constant, arbitrary initial phase that fluctuates over the many repeats of one observation (i.e., the nuclear bath remains asynchronous). Assuming as before a uniform spin density $\kappa(\vec{r}) = \kappa$ and with the correspondence $\sum_{j=1}^{M} \to \int_{Nuclear\ ensemble} d^3r\,\kappa$ (see Eq.(15)) one finds after ensemble average

$$\left[S_Q^{HE}(\tau)\right]_{Order} \sim 1 - \sin^4\left(\frac{\omega_{N\_0}\tau}{2}\right)\left(\frac{4\pi\kappa C^2}{3r_{min}^3} + \left(3\pi T_N \kappa C \ln\left(\frac{r_{max}}{r_{min}}\right)\right)^2 - \frac{4\pi\kappa C^2 T_N^2}{r_{min}^3}\right), \quad (18)$$

where the subscript emphasizes the special case of assumed nuclear spin order. Expanding (18) to get the sum of independent terms, we identify two types of contributions: The second and fourth terms (originating from the first and third sums in (17), respectively) are 'short range', in the sense that they become negligible when $r_{min}$ exceeds a few nanometers, as discussed above (Eq. (16)). The third term, however, is 'long-range' as it grows logarithmically with the size of the nuclear ensemble characterized by $r_{max}$ (assumed much greater than $r_{min}$).

As the nuclear spin order can be simply specified as a well-chosen set of $\varepsilon_j(\theta,\varphi)$, we use the semi-classical formulae derived above to help us interpret the meaning of this contribution: Starting from the upper half of Eq. (14) and assuming the same relative phases $\varepsilon_j(\theta,\varphi)$, we find the expression

$$\left[S_C^{HE}(\tau)\right]_{Order} \sim 1 - \sin^4\left(\frac{\omega_{N\_0}\tau}{2}\right)\left(3\pi T_N \kappa C \ln\left(\frac{r_{max}}{r_{min}}\right)\right)^2, \tag{19}$$

implying that $\left[S_Q^{HE}(\tau)\right]_{Order} \to \left[S_C^{HE}(\tau)\right]_{Order}$ when $r_{min}$ is sufficiently large.

While Eqs. (18) and (19) are strictly valid only for our chosen *ordered* nuclear spin configuration, generally spin order leads to qualitatively-different, non-local contributions that may ultimately dominate the signal response. Changing from 'short-' to 'long-range' can be loosely characterized by a critical value $r_{crit}$ dependent on the specific geometry of the problem and defined by the conditions

$$\sum_{j=1}^{M}\left(T_N^{(j)}\tilde{\theta}_j \sin(\omega_{N\_0}\tau - \tilde{\varphi}_j - \varepsilon_j)\right)^2 < \left(\sum_{j=1}^{M} T_N^{(j)}\tilde{\theta}_j \sin(\omega_{N\_0}\tau - \tilde{\varphi}_j - \varepsilon_j)\right)^2$$

$$\sum_{j=1}^{M}\left(T_N^{(j)}\tilde{\theta}_j \sin(\omega_{N\_0}\tau - \tilde{\varphi}_j - \varepsilon_j)\right)^2 \ll 1 \tag{20}$$

We conclude, therefore, that nuclear spin order marks the emergence of 'long-range' fields and leads to 'classical' spin magnetometry (Eq. (13)) when all contributing nuclear spins lie beyond $r_{crit}$.

For completeness, we mention that similar mechanisms are at work in the case of a Ramsey sequence (see Appendix C). In analogy to Eq. (11), one can derive a quantum mechanical expression for the Ramsey signal $S_Q^R(t)$ that explicitly takes into account the longitudinal and transverse couplings with the surrounding nuclear spin bath (Eq. (C3)). In the absence of a refocusing pulse, $S_Q^R(t)$ decays monotonically over a time $T_2^*$ (see Eq. (C4)) in a way that, nonetheless, does not depend on the rms amplitude of the nuclear field fluctuations (compare with Eq. (4)). Rather than a superposition of signals with slightly different frequencies — our intuitive, classical interpretation of the signal decay — the NV response falls off as a consequence of quantum interference arising from differences in the nuclear spin evolution introduced by the NV center itself (operators $U_{00}$ and $U_{11}$ in Eq. (C1)). A practical outcome is that one cannot 'burn a hole' in the magnetic resonance spectrum of a single NV center: A 100 μs long, very low power pulse collapses one entire peak of the $^{14}$N triplet. Similar to the Hahn-echo case, the Ramsey decay is caused by short-range interactions and has a quantum (i.e., non-classical) origin. Finally, it is not difficult to prove that a classical description of the interplay between the NV center and surrounding nuclear spins re-emerges in the presence of nuclear spin order.

III.d. Quantum Approach to the Correlation Protocol

We now tackle the pulse protocol of Fig. 2(a) used for measuring correlations of the magnetic field. Based on the expressions derived for the Hahn-echo protocol, one expects non-negligible signatures from the 'spin noise' terms in the second sum of Eq. (11) when the experiment is designed to detect long-term correlations of the nuclear bath. Unfortunately, and unlike the Ramsey or Hahn-echo experiments, the pulse sequence is complex enough to make the derivation of analytical formulae impractical. We can, nonetheless, estimate the amplitude of such fluctuations. For this purpose we consider the second sum in Eq. (11)

$$V^{HE} \equiv 8\sin^2(\omega_{N\_0}\tau/2)\sum_{\substack{j,k=1\\j\neq k}}^{M} T_N^{(j)}T_N^{(k)}\widetilde{\theta}_j\widetilde{\theta}_k \sin(\omega_{N\_1}^{(j)}\tau/2)\sin(\omega_{N\_1}^{(k)}\tau/2)\times$$

$$\times \sin\left((\omega_{N\_0}+\omega_{N\_1}^{(j)})\tau/2-\widetilde{\varphi}_j-\varepsilon_j\right)\sin\left((\omega_{N\_0}+\omega_{N\_1}^{(k)})\tau/2-\widetilde{\varphi}_k-\varepsilon_k\right) \quad (21)$$

$$\approx 8\sin^4(\omega_{N\_0}\tau/2)\sum_{\substack{j,k=1\\j\neq k}}^{M} T_N^{(j)}T_N^{(k)}\widetilde{\theta}_j\widetilde{\theta}_k \sin(\omega_{N\_0}\tau-\widetilde{\varphi}_j-\varepsilon_j)\sin(\omega_{N\_0}\tau-\widetilde{\varphi}_k-\varepsilon_k),$$

and calculate the rms amplitude $\sigma_V^{HE} = \sqrt{\langle V^2 \rangle}$. Retaining terms to second order in $\widetilde{\theta}$, a lengthy but straightforward calculation yields

$$\sigma_V^{HE}(\tau) \approx 8\sin^4\left(\frac{\omega_{N\_0}\tau}{2}\right)T_N^2\sqrt{\frac{3}{4}\sum_{\substack{j,k=1\\j\neq k}}^{M}\widetilde{\theta}_j^2\widetilde{\theta}_k^2}, \quad (22)$$

where we assumed for simplicity $\langle (T_N^{(j)})^2 \rangle = T_N^2$ for all spins $j$. By comparison to Eq. (12), we conclude that when a collapse takes place ($\omega_{N\_0}\tau \sim \pi$), bath fluctuations introduce changes in the NV signal as large as the depth of the collapse itself.

To better understand the role of these fluctuations in the correlation protocol of Fig. 2(a) we conducted a numerical simulation that takes into account interactions with hundreds of $^{13}$C spins randomly distributed over a virtual diamond lattice (Fig. 4(a)). To bring the computing time down to realistic values, we followed a 'disjoint cluster' approach where dipolar couplings between distinct nuclear spins in the lattice are taken into account only for clusters of nuclear spins, up to a predefined cluster size threshold.[21] Similar to Fig. 3(a), $S_Q^{Cor}(\tau=\pi/\omega_{N\_0},\widetilde{\tau})$ exhibits periodic maxima (or minima) when $\widetilde{\tau}=2m\pi/\omega_{N\_0}$ (or $\widetilde{\tau}=(2m+1)\pi/\omega_{N\_0}$). To more clearly expose the influence of homonuclear dipolar couplings on the overall pattern, we conducted an analogous computation that ignored the effect of $H_N'$ altogether. As expected, we find that $^{13}$C-$^{13}$C interactions are only responsible for a slow decay over a time $T_{2\_N}$ and can, as before, be ignored when describing the main signal features.

Despite the absence of analytical formulae, one can interpret these results within the framework of our prior discussion. We start by noting that $S_Q^{Cor}$ depends on the $^{13}$C spatial distribution and markedly changes from one NV to the next (Figs. 4(a) and 4(b)). Overall, however, $S_Q^{Cor}$ never exceeds a few percents of the allowed range (from −1 to 1), a value below our observation limit (of order ~10 %, see Fig. 3(b)) and thus undetectable in our present experimental conditions. In agreement with Eq. (22), this small signal amplitude is not indicative of weak nuclear fields. Rather than a smooth, sinusoidal oscillation, $S_Q^{Cor}$ features sharp crests and valleys implying that the NV response is far from the linear regime (see Fig. 2 and main narrative in Section II). We surmise, therefore, that the null response in Fig. 3b is truly the result of a weak correlation between the nuclear spin bath configurations at the beginning of the first and second encoding segments of the detection protocol ($t = 0$ and $t = 2\tau + \tilde{\tau}$, respectively). The latter, in turn, seems to be a consequence of the NV center entanglement with neighboring nuclear spins (an effect we gauged above via the second sum in Eq. (11)); i.e., the very encoding process alters the bath configuration and thus diminishes the correlation between two consecutive observations. Similar to Eq. (18) (and Eq. (C5) in Appendix C) we speculate that the classical response (Fig. 3(a)) resurfaces when the state of individual nuclear spins correlates, at least partly, with that of their neighbors, and when all spins are located beyond $r_{crit}$. The experiment details of this transition will likely be the subject of further work.

**Conclusion**

The use of NV centers as a platform toward mesoscopic nuclear spin magnetometry relies on detection protocols designed to interrogate large numbers of nuclear spins weakly coupled to the center. Throughout the present study we investigated the effect of these couplings on the response of NV centers to common pulse sequences, namely the Ramsey and Hahn-echo protocols. We found strong phenomenological similarities with the case where, rather than interacting with neighboring nuclear spins, the NV center evolves in the presence of a classical magnetic field designed to emulate nuclear spin bath fluctuations.

In the limit where nuclear spins are independent from each other, a closer examination reveals that these similarities are only superficial, and that fundamentally different mechanisms are at work. For the particular case of the Hahn-echo sequence, our analysis suggests that the pattern of collapses and revivals must persist, for example, if the nuclear bath is initialized to the fully polarized state. Contrary to widespread perception, we also find that field inhomogeneities produced over time by a fluctuating nuclear spin environment are not the only cause for the signal decay in a Ramsey protocol. Instead, our analysis indicates that both collapses and revivals and Ramsey decoherence are inherently quantum mechanical in nature: Rather than a decay due to random phase accumulation, the NV center becomes non-observable due to entanglement with the nuclear spin bath. This process results from anisotropic interactions effectively constrained to a small, nanometer-diameter volume and will thus be negligible if probed nuclear spins are sufficiently removed from the center. In this limit, we find that the combined NV-center/nuclear-spin system can be described well by semi-classical equations when the bath exhibits long-range order.

Even though both semi-classical and quantum descriptions ultimately predict identical Hahn-echo or Ramsey patterns, the seemingly subtle underlying differences have a dramatic effect in more general scenarios. We illustrated the problem through a pulse sequence designed to provide information on the long-term correlations of a fluctuating magnetic field acting on the NV center. While our approach succeeds in exposing the central frequency, rms amplitude and correlation time of a classical, random magnetic field, it fails to extract the same information when we consider nuclear spins close to the center as the source of the field. We interpret our observations as an unwelcome but unavoidable consequence of the interrogation process itself, whose influence on the bath evolution alters the time correlation it would otherwise exhibit.

Disjoint cluster simulations of the integrated probe-bath system suggest that not all information is lost, and that the nuclear spin correlation pattern should be observable if higher signal-to-noise ratio conditions are reached. In this regard, one possibility could be, for example, the use of diamond crystals configured in the form of a solid-immersion lens, where much higher NV fluorescence count rates have been reported.[22,23] Such observations would shed light on the nuclear bath dynamics without the limitations imposed by NV decoherence. In particular, one could, for example, tailor the protocol to directly gauge $^{13}$C spin relaxation caused by varying NV illumination during the free-evolution time $\tilde{\tau}$.[24] Finally, and while the initial state of the bath considered here only takes into account single spin coherences (Eq. (10)), one can envision modified sequences where stochastic multiple-quantum coherences of the nuclear system are systematically converted into single-quantum prior to observation. This class of schemes could be useful, for instance, to provide information on the mesoscale distribution of $^{13}$C spins (whose relatively weak couplings do not lead to resolved splittings in the NV spectrum).

**Acknowledgements**

We thank Prof. Glen Kowach for providing the diamond sample. Some of the derivations were motivated by helpful discussions with Profs. Lilian Childress and David Cory. We acknowledge support from Research Corporation and from the National Science Foundation under project CHE-0545461.

**Appendix A: Derivation of Eq. (5)**

Starting from Eq. (1) and using $\rho_{NV}(0) = |0\rangle\langle 0| = (1/2)(1+\sigma_z)$ we find at the end of the correlation protocol $(\pi/2)_y - \tau - (\pi)_y - (\pi/2)_x - \tilde{\tau} - (\pi/2)_y - \tau - (\pi)_y - (\pi/2)_x$

$$\rho_{NV}(2\tau+\tilde{\tau}) = \frac{1}{2}(1 + \sigma_y \cos\phi_{12}\cos\phi_3 - \sigma_x(\sin\phi_{12}\cos\phi_{45} + \sin\phi_3 \cos\phi_{12}\sin\phi_{45}) + \quad (A.1)$$
$$+ \sigma_z(\sin\phi_{12}\sin\phi_{45} - \sin\phi_3 \cos\phi_{12}\cos\phi_{45})),$$

where we used the notation in the main text and $\phi_3 = (2\gamma_{NV}b_{A\_1}/\omega_A)\sin(\omega_A\tilde{\tau}/2)$ $\cos(\omega_A\tau + \omega_A\tilde{\tau}/2 + \varepsilon_A^{(3)})$ with $\varepsilon_A^{(3)}$ denoting the magnetic field phase at the beginning of the $\tilde{\tau}$ interval; $\sigma_x, \sigma_y, \sigma_z$ denote, as usual, the set of Pauli matrices. On the other hand, the modified protocol $(\pi/2)_y - \tau - (\pi)_x - (\pi/2)_{-x} - \tilde{\tau} - (\pi/2)_y - \tau - (\pi)_y - (\pi/2)_x$ yields

$$\rho_{NV}(2\tau+\tilde{\tau}) = \frac{1}{2}(1 - \sigma_y \cos\phi_{12}\cos\phi_3 + \sigma_x(-\sin\phi_{12}\cos\phi_{45} + \sin\phi_3 \cos\phi_{12}\sin\phi_{45}) + \quad (A.2)$$
$$+ \sigma_z(\sin\phi_{12}\sin\phi_{45} + \sin\phi_3 \cos\phi_{12}\cos\phi_{45})).$$

Therefore, if the proper phase cycling is introduced from one repeat to the next, we obtain

$$S_C^{Cor}(\tau, \tilde{\tau}) = \langle \sin\phi_{12}\sin\phi_{45} \rangle \quad (A.3)$$

in agreement with Eq. (5). For completeness, we mention that for the conditions of the present experiment we find (both experimentally and numerically) that $\langle \sin\phi_3 \cos\phi_{12}\cos\phi_{45} \rangle \sim 0$. In practice, this makes the phase cycling unnecessary. Also, we note that other correlation functions (including $\langle \cos\phi_{12}\cos\phi_{45} \rangle$, $\langle \sin\phi_{12}\cos\phi_{45} \rangle$, etc) can be obtained with a proper selection of the relative phases within the same general pulse protocol.

**Appendix B: Derivation of equation (11)**

For presentation purposes, we rewrite Eq. (10) in the form $S_Q^{HE}(\tau) = \left\langle \text{Re}\left\{ \prod_{j=1}^{M} S_j^{HE} \right\} \right\rangle$

where we defined

$$S_j^{HE} \equiv Tr_r\{\rho_N^{(j)}(U^\dagger)_{00}(U^\dagger)_{11}(U)_{00}(U)_{11}\} =$$
$$= Tr_r\{\rho_N^{(j)} e^{i\omega_{N\_0} I_z^{(j)}\tau} e^{i\vec{\omega}_{N\_1}^{(j)}\cdot\vec{I}^{(j)}\tau} e^{-i\omega_{N\_0} I_z^{(j)}\tau} e^{-i\vec{\omega}_{N\_1}^{(j)}\cdot\vec{I}^{(j)}\tau}\}.$$ (B1)

After some algebra and using $|\pm\rangle$ to denote the two eigenfunctions of $I_z$ (nuclei are assumed to have spin $I=1/2$) one can show that

$$\langle+|e^{i\vec{\omega}_{N\_1}^{(j)}\cdot\vec{I}^{(j)}\tau}|+\rangle = \left(\langle-|e^{i\vec{\omega}_{N\_1}^{(j)}\cdot\vec{I}^{(j)}\tau}|-\rangle\right)^* = \cos^2\left(\frac{\tilde{\theta}_j}{2}\right)e^{i\omega_{N\_1}^{(j)}\tau/2} + \sin^2\left(\frac{\tilde{\theta}_j}{2}\right)e^{-i\omega_{N\_1}^{(j)}\tau/2}$$

$$\langle+|e^{i\vec{\omega}_{N\_1}^{(j)}\cdot\vec{I}^{(j)}\tau}|-\rangle = -\left(\langle-|e^{i\vec{\omega}_{N\_1}^{(j)}\cdot\vec{I}^{(j)}\tau}|+\rangle\right)^* = i\sin\tilde{\theta}_j \sin\left(\frac{\omega_{N\_1}^{(j)}\tau}{2}\right)e^{-i\tilde{\varphi}_j}$$ (B2)

and from here we get

$$S_j^{HE} = 1 - 2\sin^2\tilde{\theta}_j \sin^2\left(\frac{\omega_{N\_1}^{(j)}\tau}{2}\right)\sin^2\left(\frac{\omega_{N\_0}\tau}{2}\right) + iP_N^{(j)}\sin^2\tilde{\theta}_j \sin^2\left(\frac{\omega_{N\_1}^{(j)}\tau}{2}\right)\sin(\omega_{N\_0}\tau) -$$
$$- 4iT_N^{(j)}\sin\tilde{\theta}_j \sin\left(\frac{\omega_{N\_1}^{(j)}\tau}{2}\right)\sin\left(\frac{\omega_{N\_0}\tau}{2}\right)\left(\cos^2\left(\frac{\tilde{\theta}_j}{2}\right)\sin\left((\omega_{N\_0}+\omega_{N\_1}^{(j)})\frac{\tau}{2}-\tilde{\varphi}_j-\varepsilon_j\right) +$$
$$+ \sin^2\left(\frac{\tilde{\theta}_j}{2}\right)\sin\left((\omega_{N\_0}-\omega_{N\_1}^{(j)})\frac{\tau}{2}-\tilde{\varphi}_j-\varepsilon_j\right)\right)$$ (B3)

For $\tilde{\theta}_j \ll 1$ B3 takes the form

$$S_j^{HE} = 1 - 2\tilde{\theta}_j^2 \sin^2\left(\frac{\omega_{N\_1}^{(j)}\tau}{2}\right)\sin^2\left(\frac{\omega_{N\_0}\tau}{2}\right) + iP_N^{(j)}\tilde{\theta}_j^2 \sin^2\left(\frac{\omega_{N\_1}^{(j)}\tau}{2}\right)\sin(\omega_{N\_0}\tau) -$$
$$- 4iT_N^{(j)}\tilde{\theta}_j \sin\left(\frac{\omega_{N\_1}^{(j)}\tau}{2}\right)\sin\left(\frac{\omega_{N\_0}\tau}{2}\right)\sin\left((\omega_{N\_0}+\omega_{N\_1}^{(j)})\frac{\tau}{2}-\tilde{\varphi}_j-\varepsilon_j\right)$$ (B4)

Noting that $S_j^{HE}$ has the form $S_j^{HE} = (1-a_j) + ib_j$ with $a_j, b_j < 1$, we find

$$\text{Re}\left\{\prod_{j=1}^M S_j^{HE}\right\} \approx (1-a_1)...(1-a_M) - b_1 b_2(1-a_3)...(1-a_M) - (1-a_1)b_2 b_3(1-a_3)...(1-a_M) -$$
$$... - (1-a_1)...(1-a_l)b_l(1-a_{l+1})...(1-a_r)b_r(1-a_{r+1})...(1-a_M) - ... + \text{higher order terms},$$ (B5)

which leads to

$$S_Q^{HE}(\tau) \cong \left\langle 1 - 2\sin^2(\omega_{N\_0}\tau/2)\left\{\sum_{j=1}^{M}\tilde{\theta}_j^2 \sin^2(\omega_{N\_1}^{(j)}\tau/2) + 4\sum_{\substack{j,k=1 \\ j\neq k}}^{M} T_N^{(j)}T_N^{(k)}\tilde{\theta}_k\tilde{\theta}_j \sin(\omega_{N\_1}^{(j)}\tau/2)\sin(\omega_{N\_1}^{(k)}\tau/2) \times \right.\right.$$

$$\left.\left. \times \sin\left((\omega_{N\_0}+\omega_{N\_1}^{(j)})\tau/2 - \tilde{\varphi}_j - \varepsilon_j\right)\sin\left((\omega_{N\_0}+\omega_{N\_1}^{(k)})\tau/2 - \tilde{\varphi}_k - \varepsilon_k\right)\right\}\right\rangle.$$

(B6)

**Appendix C: Ramsey sequence**

Starting from Eq. (9) we find for a Ramsey sequence

$$S_Q^R(t) = \left\langle \mathrm{Re}\left\{\prod_{j}^{M} Tr_r\left\{(U^\dagger)_{00}(U)_{11}\rho_N^{(j)}\right\}\right\}\right\rangle$$

(C1)

where $U$ is the free evolution operator over time $t$. Using $\rho_N^{(j)} = 1/2 + 2P_N^{(j)}I_z^{(j)} + 2T_N^{(j)}I_x^{(j)}\cos\varepsilon_j + 2T_N^{(j)}I_y^{(j)}\sin\varepsilon_j$, we obtain

$$Tr_r\left\{(U^\dagger)_{00}(U)_{11}\rho_N^{(j)}\right\} = \cos^2\left(\frac{\tilde{\theta}}{2}\right)\cos\left((\omega_{N\_0}-\omega_{N\_1}^{(j)})\frac{t}{2}\right) + \sin^2\left(\frac{\tilde{\theta}}{2}\right)\cos\left((\omega_{N\_0}+\omega_{N\_1}^{(j)})\frac{t}{2}\right) +$$

$$+ 2iP_N^{(j)}\left[\cos^2\left(\frac{\tilde{\theta}}{2}\right)\sin\left((\omega_{N\_0}-\omega_{N\_1}^{(j)})\frac{t}{2}\right) + \sin^2\left(\frac{\tilde{\theta}}{2}\right)\sin\left((\omega_{N\_0}+\omega_{N\_1}^{(j)})\frac{t}{2}\right)\right] -$$

$$- 2iT_N^{(j)}\sin\tilde{\theta}_j\sin\left(\frac{\omega_{N\_1}^{(j)}t}{2}\right)\cos\left(\frac{\omega_{N\_0}t}{2} - \tilde{\varphi}_j - \varepsilon_j\right).$$

(C2)

After replacing in (C1) and in the approximation discussed in Appendix B, we find

$$S_Q^R = \left\langle A - \frac{1}{2}\sin\left(\frac{\omega_{N\_0}t}{2}\right)\sum_{j=1}^{M}B_j\tilde{\theta}_j^2\sin\left(\frac{\omega_{N\_1}^{(j)}t}{2}\right) - 2\sum_{\substack{j,k \\ j\neq k}}^{M}C_{jk}\left(P_N^{(j)}P_N^{(k)}\Gamma_{jk} + T_N^{(j)}T_N^{(k)}\mathrm{K}_{jk} - 2T_N^{(j)}P_N^{(k)}\Lambda_{jk}\right)\right\rangle$$

(C3)

where

$$A = \prod_{j=1}^{M}\cos\left((\omega_{N\_0}-\omega_{N\_1}^{(j)})\frac{t}{2}\right); \quad B_j = \prod_{\substack{k=1 \\ k\neq j}}^{M}\cos\left((\omega_{N\_0}-\omega_{N\_1}^{(k)})\frac{t}{2}\right); \quad C_{jk} = \prod_{\substack{l=1 \\ l\neq j,k}}^{M}\cos\left((\omega_{N\_0}-\omega_{N\_1}^{(l)})\frac{t}{2}\right);$$

and

$$\Gamma_{jk} = \left(1-\frac{\widetilde{\theta}_j^2}{2}\right)\sin\left((\omega_{N\_0}-\omega_{N\_1}^{(j)})\frac{t}{2}\right)\sin\left((\omega_{N\_0}-\omega_{N\_1}^{(k)})\frac{t}{2}\right) + \frac{\widetilde{\theta}_j^2}{2}\sin\left((\omega_{N\_0}+\omega_{N\_1}^{(j)})\frac{t}{2}\right)\sin\left((\omega_{N\_0}-\omega_{N\_1}^{(k)})\frac{t}{2}\right)$$

$$\mathrm{K}_{jk} = \widetilde{\theta}_j\widetilde{\theta}_k \sin\left(\frac{\omega_{N\_1}^{(j)}t}{2}\right)\sin\left(\frac{\omega_{N\_1}^{(k)}t}{2}\right)\cos\left(\frac{\omega_{N\_0}t}{2}-\widetilde{\varphi}_j-\varepsilon_j\right)\cos\left(\frac{\omega_{N\_0}t}{2}-\widetilde{\varphi}_k-\varepsilon_k\right)$$

$$\Lambda_{jk} = \widetilde{\theta}_j \sin\left(\frac{\omega_{N\_1}^{(j)}t}{2}\right)\cos\left(\frac{\omega_{N\_0}t}{2}-\widetilde{\varphi}_j-\varepsilon_j\right)\sin\left((\omega_{N\_0}-\omega_{N\_1}^{(k)})\frac{t}{2}\right).$$

Assuming independent, uncorrelated nuclear spins and after ensemble average, $\Gamma_{jk}$, $\mathrm{K}_{jk}$, and $\Lambda_{jk}$ cancel and we can rewrite (C3) in the simpler form

$$S_Q^R = A - \frac{1}{2}\sin\left(\frac{\omega_{N\_0}t}{2}\right)\sum_{j=1}^M B_j \widetilde{\theta}_j^2 \sin\left(\frac{\omega_{N\_1}^{(j)}t}{2}\right) \sim \exp\left(-\left(\frac{t}{T_2^*}\right)^2\right)\left(1-\frac{1}{2}\sin\left(\frac{\omega_{N\_0}t}{2}\right)\sum_{j=1}^M \widetilde{\theta}_j^2 \sin\left(\frac{\omega_{N\_1}^{(j)}t}{2}\right)\right), \quad (C4)$$

where $T_2^*$ is a parameter characterizing the time decay resulting from $A$ and $B_j$.

When spin order is present and assuming $r_{min} > r_{crit}$ (see main text), $\sum_{\substack{j,k=1 \\ j\neq k}}^M \to \left(\sum_{j=1}^M\right)^2$ and Eq. (C3) can be cast in the form

$$S_Q^R \approx \exp\left(-\left(\frac{t}{T_2^*}\right)^2\right)\left[1-\frac{1}{2}\left\langle\left(\sum_{j=1}^M P_N^{(j)}A_{zz}t\right)^2\right\rangle - \sin^2\left(\frac{\omega_{N\_0}t}{2}\right)\left\langle\left(\sum_{j=1}^M T_N^{(j)}\widetilde{\theta}_j \cos\left(\frac{\omega_{N\_0}t}{2}-\varepsilon_j-\varphi_j\right)\right)^2\right\rangle\right]. \quad (C5)$$

The above expression coincides with Eq. (4) in the main text if we make use of the correspondence with the field produced by classical magnetic moments

$$\gamma_{NV} b_{N\_0} \to \sum_{j=1}^M P_N^{(j)} A_{zz}^{(j)}$$

$$\gamma_{NV} b_{N\_1} \cos(\omega_{N\_0}t+\varepsilon_N) \to \sum_{j=1}^M T_N^{(j)}\left(A_{zx}^{(j)}\cos(\omega_{N\_0}t+\varepsilon_j) + A_{zy}^{(j)}\sin(\omega_{N\_0}t+\varepsilon_j)\right) \quad (C6)$$

$$\cong \omega_{N\_0}\sum_{j=1}^M T_N^{(j)}\widetilde{\theta}_j \cos(\omega_{N\_0}t+\varepsilon_j-\widetilde{\varphi}_j),$$

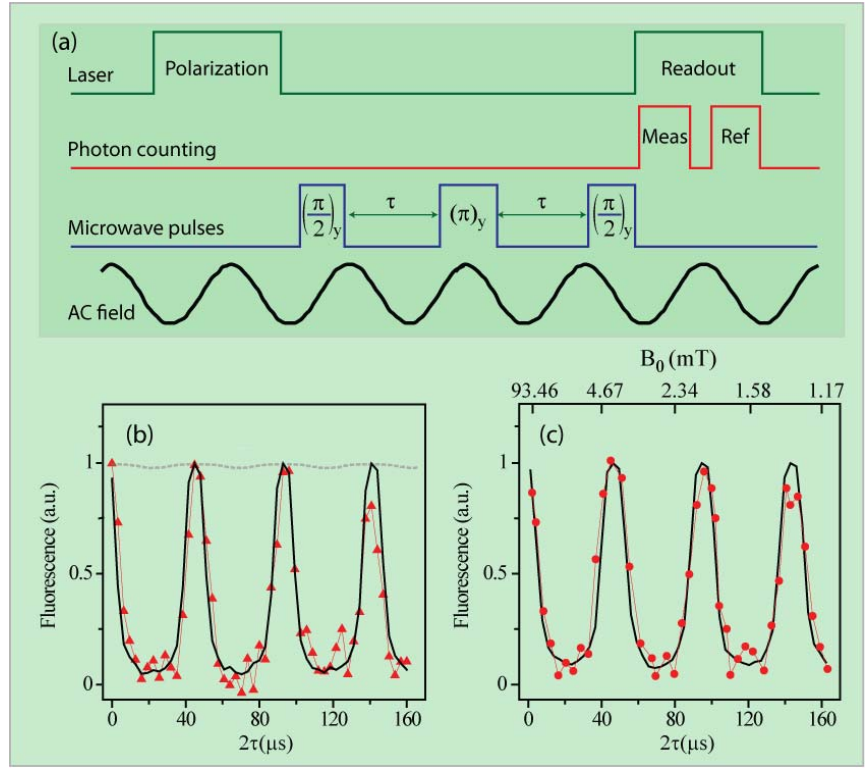

**Fig. 1**: (a) Schematics of the detection protocol in a Hahn-echo sequence. An AC field of variable amplitude, phase and frequency is used when necessary as described in the text. (b) $^{13}$C-induced pattern of collapses and revivals for $B_0$=4 mT. In this case no AC field is applied. The solid line is a fit to the experimentally observed response (triangles) using Eq. (2) with $\sqrt{\langle b_{N\_1}^2 \rangle}$=4.64 µT and $\omega_N/2\pi$=42.5 kHz. The fainted, dashed line corresponds to Eq. (2) for $\sqrt{\langle b_{N\_1}^2 \rangle}$=0.1 µT (see main text). (c) Hahn-echo sequence in the presence of an asynchronous AC field of rms amplitude $\sqrt{\langle b_{A\_1}^2 \rangle}$=3.2 µT and frequency $\omega_A/2\pi$=44.5 kHz. The upper horizontal axis indicates the value of $B_0$ at each time $\tau$. Good agreement is found between experiment (circles) and Eq. (3) (solid lines).

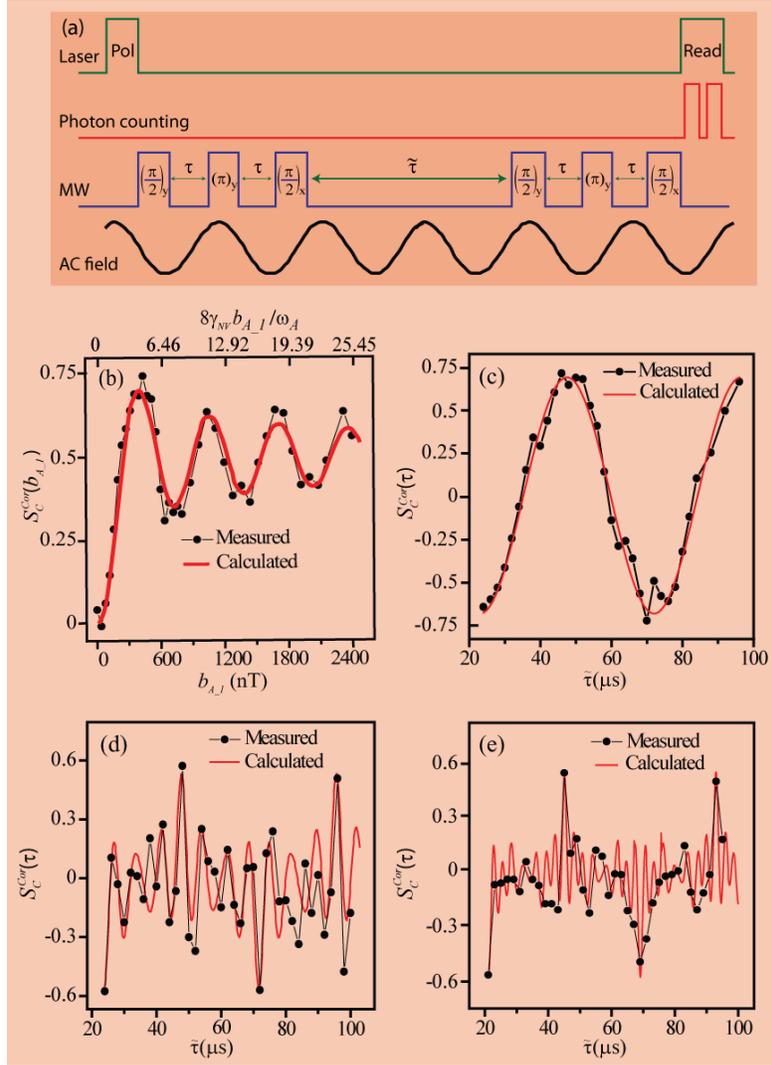

**Fig. 2:** (a) Correlation protocol in the presence of an asynchronous AC field. (b) NV response as a function of AC field amplitude $b_{A\_1}$ for the case $\omega_A(2\tau + \tilde{\tau}) = 2m\pi$. Here we choose $\tilde{\tau} = 2\tau = 48$ μs. (c-e) Signal as a function of $\tilde{\tau}$ for $b_{A\_1}$=440 nT, 2.2 μT, and 4.4 μT, respectively. In (b) through (e), circles represent experimental points and solid curves are calculated using Eq. (5). In all cases, we use $\omega_A/2\pi$=20.8 kHz so that $2\omega_A\tau = 2\pi$. For the present time scale, we virtually eliminate the effect of $^{13}$C spins by setting the static field $B_0$=3.9 mT so that $\omega_N\tau = 2\pi$.

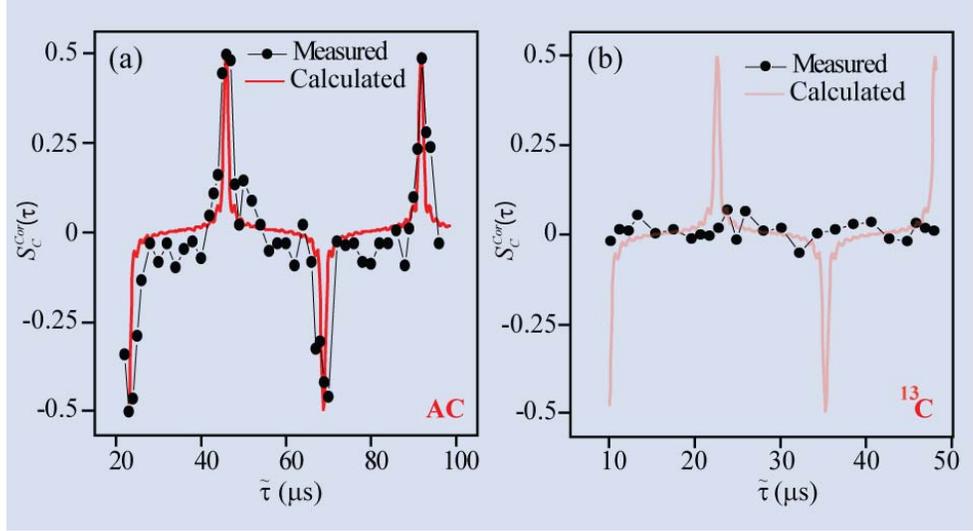

**Fig. 3:** (a) Correlation signal as a function of $\tilde{\tau}$ for an asynchronous AC field of rms amplitude $\sqrt{\langle b_{A\_1}^2 \rangle}$ =4.4 µT and frequency $\omega_A/2\pi$ =21.7 kHz. We set $\tau$=23 µs for $B_0$=4 mT (b) Analog protocol adapted to $^{13}$C spins. Maintaining $B_0$=4 mT we set $\tau$ =11.5 µs so that $2\omega_N\tau = 2\pi$. The fainted solid line is the expected pattern assuming $\sqrt{\langle b_{N\_1}^2 \rangle}$ =4.4 µT (see Fig. 1(a)) and frequency $\omega_N/2\pi$ = 43.5 kHz.

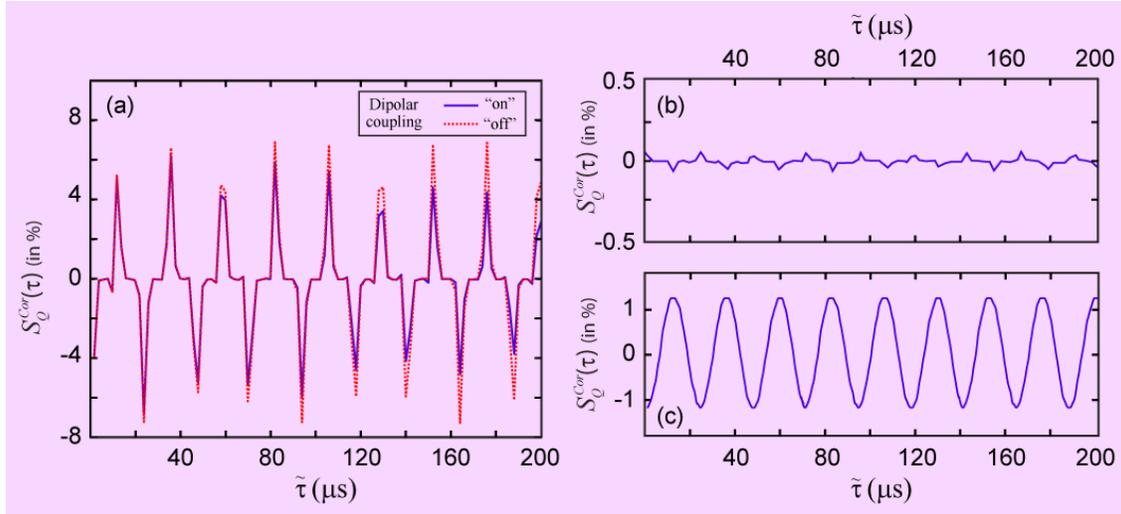

**Fig. 4:** (a) "Disjoint cluster" simulation of the NV response to the correlation protocol of Fig. 2a. Whether or not homonuclear dipolar couplings are included in the simulation (solid and dashed lines, respectively), the calculated pattern resembles that in Fig. 3a, though of much smaller amplitude. For these calculations we considered a total of 834 $^{13}$C nuclei randomly distributed over a 21x21x21 unit cell diamond lattice. (b) Simulations over different NV centers display strong signal variability. Calculations over diverse $^{13}$C environments indicate, however, that $S_Q^{Cor}$ never exceeds a few percent of the allowed maximum. (c) As expected, $S_Q^{Cor}$ transitions to a sinusoidal shape when $^{13}$C spins (totaling ~840 over the simulated crystal) are restricted to more than ten unit cells away from the NV center.